\documentclass[12pt]{article}
\usepackage{a4wide}
\usepackage{epsfig}

\newcommand{\slk}{/\kern-6pt k}
\newcommand{\sll}{/\kern-4pt l}
\newcommand{\slp}{p\kern-5pt/}
\newcommand{\slq}{q\kern-5.5pt/}
\newcommand{\sls}{s\kern-5.5pt/}
\newcommand{\tr}{\mathop{\rm tr}\nolimits}

\newcommand{\bea}{\begin{eqnarray}}
\newcommand{\ena}{\end{eqnarray}}
\newcommand{\nn}{\nonumber\\}
\newcommand{\be}{\begin{equation}}
\newcommand{\en}{\end{equation}}
\newcommand{\oone}{\hbox{$1\kern-2.5pt\hbox{\rm l}$}}
\newcommand{\ssigma}{\hbox{$\kern2.5pt\vrule height4pt\kern-2.5pt\sigma$}}

\newcommand{\GeV}{{\rm\,GeV}}

\newcommand\pfrac[2]{\left(\frac{#1}{#2}\right)}
\newcommand{\Li}{{\rm Li}}

\newcommand{\IP}{\mbox{I}\!\mbox{P}}

\newcommand{\slell}{/\kern-5pt\ell}

\begin{document}

\thispagestyle{empty} 
\begin{flushright}
Alberta Thy 3-18\\
MITP/18-008\\
SI-HEP-2018-11\\
\end{flushright}
\vspace{0.5cm}

\begin{center}

{\Large\bf NNLO QCD corrections to the polarized top quark decay
    \boldmath$t(\uparrow) \to X_b+W^+$}\\[1.3cm]
{\large A.~Czarnecki$^1$,  S.~Groote$^2$, J.G.~K\"orner$^3$, 
and J.H.~Piclum$^4$}\\[1cm]
$^1$ Department of Physics, University of Alberta, Edmonton,
Alberta T6G 2E1, Canada\\[7pt]
$^2$ Loodus- ja t\"appisteaduste valdkond, F\"u\"usika Instituut,\\[.2cm]
  Tartu \"Ulikool, W.~Ostwaldi 1, 50411 Tartu, Estonia\\[7pt]
$^3$ Institut f\"ur Physik, Johannes-Gutenberg-Universit\"at,\\[.2cm]
  Staudinger Weg 7, 55099 Mainz, Germany\\[7pt]
$^4$ Theoretische Physik 1, Naturwissenschaftliche-Technische Fakult\"at,
  \\[.2cm]
Universit\"at Siegen, 57068 Siegen, Germany
\end{center}

\vspace{1cm}
\begin{abstract}\noindent
We compute the next-to-next-to-leading order (NNLO) QCD corrections to
the decay $t(\uparrow) \to X_b +W^+$ of a polarized top quark. The
spin-momentum correlation in this quasi two-body decay is described by
the polar angle distribution $\mathrm{d}\Gamma/\mathrm{d}\cos\theta_P
=\frac{\Gamma}{2}(1+P_t\, \alpha_P\, \cos\theta_P)$ where $P_t$ is
the polarization of the top quark and $\alpha_P$ denotes the asymmetry
parameter of the decay. For the latter we find
$\alpha^{\mathrm{NNLO}}_P=0.3792\pm 0.0037$.
\end{abstract}

\newpage

%%%%%%%%%%%%%%%%%%%%%%%%%%%%%%%%%%%%%%%%%%%%%%%%%%%%%%%%%%%%%%%%%%%%%%%%%%%%
%%%%%%%%%%%%%%%%%%%%%%%%%%%%%%%%%%%%%%%%%%%%%%%%%%%%%%%%%%%%%%%%%%%%%%%%%%%%%
\section{Introduction}
The number of single top quark events reported by the LHC collaborations
ATLAS and CMS in Run 1 and 2 is ever increasing. More and more single top quark
events have been and are being seen at the
LHC~\cite{Aaboud:2016ymp,Aaboud:2017pdi,Sirunyan:2016cdg,CMS:2016xnv}.
The present situation concerning both ATLAS and CMS results on single top
production is nicely summarized in a review article by
N.~Faltermann~\cite{Faltermann:2017vry}. After Run 3 the LHC will operate in
the High Luminosity Mode with a projected total luminosity of $3{\rm\,ab}^{-1}$
which corresponds to approximately $10^9$ single top quark events. In the
dominating t-channel process, which is a weak production process, single top
quarks are produced with a large longitudinal polarization $P_t \simeq 0.9$
in the direction of the spectator jet in the top quark rest frame, and a
slightly smaller polarization of $P_t \simeq 0.8$ for antitop
quarks~\cite{Mahlon:1999gz,Tait:2000sh,Espriu:2001vj,Espriu:2002wx}.\footnote{Close
  to maximal values of the polarization of top quarks can be achieved
  with moderate tuning of the longitudinal beam polarization at the
  ILC (see e.g.\ Ref.~\cite{Groote:2010zf}).} Since the top quark
decays so rapidly, it retains its polarization from birth when it
decays. The dominant decay mode is the quasi-two-body mode
$t(\uparrow) \to X_b +W^+$ mediated by the quark level transition $t
\to b$ proportional to the CKM matrix element $V_{tb} \approx 1$.

In this paper we study top quark polarization effects in the quasi two-body
decay $t(\uparrow) \to X_b +W^+$ at next-to-next-to-leading order
(NNLO) in QCD. The NNLO results are 
obtained in the form of a power series expansion in terms of the ratio
$x=m_W/m_t$, where $m_W$ and $m_t$ are the masses of the $W$ boson and
the top quark, and we include terms up to $x^{10}$. This 
analysis can be considered to be complementary to the decay part of the recent
numerical NNLO evaluation of polarized top production and
decay~\cite{Berger:2016oht,Berger:2017zof}.

\begin{figure}[t]
\begin{center}
\epsfig{figure=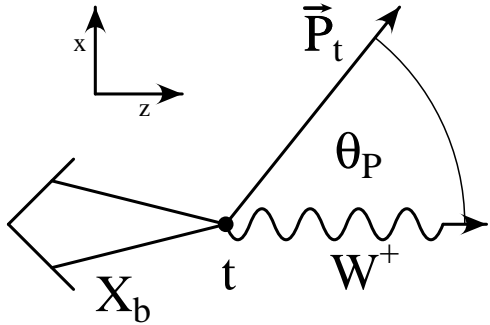, scale=0.7}
\end{center}
\caption{\label{polar}Definition of the polar angle $\theta_{P}$ in the decay 
$t(\uparrow) \to X_b + W^{+}$.}
\end{figure}

Since the decay is weak, the top quark is self-analyzing. The angular decay
distribution reads
\be\label{polar2body}
\frac1\Gamma\frac{\mathrm{d}\Gamma}{\mathrm{d}\cos\theta_P}=
\frac12(1+P_t\,\alpha_P\,\cos\theta_P),
\en
where $\theta_P$ is the angle between the polarization direction of the top
quark and the momentum direction of the $W^+$ (see Fig.~\ref{polar}). The
analyzing power for the polarization of the decay is given by the
asymmetry parameter $\alpha_P$ where, at leading order (LO), one has
$\alpha^{\mathrm{LO}}_P=(1-2x^2)/(1+2x^2)=0.398$. Here and throughout
this paper we set the bottom quark mass to zero.

The measurements suggested here require the reconstruction of the momentum
direction of the $W$ boson which is not simple experimentally. However, the
experimentalists have devised sophisticated tools to reconstruct the $W$-boson
momentum direction for their analysis of
the helicity fractions in unpolarized top quark decays which can also be used
in this analysis. 

This paper is organized as follows. In the next Section we outline the
calculational methods used to obtain our result. In
Sec.~\ref{sec::num} we provide a numerical analysis of the decay
rate and the asymmetry parameter. A summary and outlook are given in
Sec.~\ref{sec::sum}. Analytical results for the decay rate can be
found in the Appendix.

%%%%%%%%%%%%%%%%%%%%%%%%%%%%%%%%%%%%%%%%%%%%%%%%%%%%%%%%%%%%%%%%%%%%%%%%%%%%%%
\section{Calculation}
%%%%%%%%%%%%%%%%%%%%%%%%%%%%%%%%%%%%%%%%%%%%%%%%%%%%%%%%%%%%%%%%%%%%%%%%%%%%%%
Our calculation follows the approach used
in Refs.~\cite{Blokland:2004ye,Blokland:2005vq,IanPhD} for the calculation of the
total unpolarized decay rate and in Ref.~\cite{Czarnecki:2010gb} for the
so-called helicity fractions of the $W$ boson. Using the optical
theorem, we compute the top-quark decay width from the imaginary part
of self-energy diagrams,
\begin{equation}
\Gamma=\frac1{m_t}\,{\rm Im}(\Sigma_t),
\end{equation}
where $\Sigma_t$ is computed from one-particle irreducible self-energy
diagrams of the top quark. We sum over the spin degrees of freedom of
the $W$ boson, i.e.\ we do not specify its helicity components as has
been done in Ref.~\cite{Czarnecki:2010gb}. Thus, we use the unitary gauge
form for the spin sum:
\be\label{unitary}
\sum_{m=\pm,L}\varepsilon^\mu(m)\varepsilon^\nu(m)=\IP^{\mu\nu}
  =-g^{\mu\nu}+\frac{q^\mu q^\nu}{m_W^2},
\en
which enters our calculation in the numerator of the $W$-boson propagator.
Here $q$ is the momentum of the $W^+$. At LO we have $q=p_t-p_b$,
where $p_t$ and $p_b$ are the momenta of the top and bottom quark,
respectively.

It is clear that the polar angle distribution is sensitive to the
longitudinal polarization vector of the top quark,
$s_t^{\ell,\mu}$. We have
\begin{equation}
  \Sigma_t = \tr\left( (\slp_t + m_t)\,
    \sls_t^\ell\, \gamma_5\, \Sigma \right),
\end{equation}
where $i\Sigma$ is the sum of the top-quark self-energy diagrams.
In the rest frame of the top quark the polarization vector reads
$s_t^{\ell,\mu}=\,(0;0,0,1)$, i.e.\ the three-dimensional polarization
vector points into the direction of the momentum of the $W$ boson ($z$
direction in Fig.~\ref{polar}). For our calculation we require a
covariant representation of the longitudinal polarization four-vector
$s_t^\ell$ which is given by
\begin{equation}\label{polvector1} 
s_t^{\ell,\mu} =\frac1{|\vec q\,|}
  \Big(q^\mu-\frac{p_t\!\cdot\!q}{m_t^2}p_t^\mu\Big),
\end{equation}
where $|\vec q\,|=\sqrt{q_0^2-q^2}$. The polarization four-vector
$s_t^{\ell,\mu}$ can be seen to satisfy $p_t\cdot s_t^\ell=0$ and
$s_t^\ell\cdot s_t^\ell=-1$ where we use the fact that
$p_t\cdot q=m_tq_0$ in the rest system of the top quark. Just as in
the case of the helicity fractions, we find that due to the
polarization vector we have to deal with the modulus of the $W$
momentum three-vector in the denominator of the expressions for the
self-energy diagrams.

There are altogether 38 three-loop diagrams. Since we use the unitary
gauge for the $W$ boson there is no need to include Goldstone bosons
in the Feynman diagrams. For the gluons we use the covariant $R_\xi$
gauge with the spin sum $\IP^{\mu\nu}(R_\xi)=-g^{\mu\nu}+\xi\, k^\mu
k^\nu/k^2$, where $\xi$ is an arbitrary gauge parameter. We have
checked that the gauge-parameter dependence cancels in the final
result. Since we only require traces involving an even number of
$\gamma_5$ matrices, we can work with a naively anticommuting
$\gamma_5$~\cite{Kreimer:1989ke,Korner:1991sx}.

After setting the bottom-quark mass to zero, the Feynman integrals
corresponding to the top-quark self-energy diagrams depend on two
scales: the hard scale $m_t$ and the soft scale $m_W$. We then employ
the method of regions (see e.g.\ Ref.~\cite{Smirnov:2002pj}) to
construct an expansion around the limit where the ratio $x=m_W/m_t$ of
the two scales tends to zero. Here, we have to consider two regions
for each loop momentum (the loop momenta are chosen to be the momenta
of the gluons and the $W$ boson). In the so-called hard region, all
components of a loop momentum $k$ scale like the hard scale $k^\mu\sim
m_t$ for $\mu\in\{0,1,2,3\}$ and in the so-called soft region all
components scale like the soft scale $k^\mu\sim m_W$. In each region
we then expand the integrand according to the scaling of all loop
momenta. If the momentum of a gluon is soft, the corresponding loop
integral becomes scaleless and is set to zero in dimensional
regularization. We are therefore left with only two contributions for
each integral: one where all loop momenta are hard and one where the
gluon momenta are hard, but the momentum of the $W$ boson is soft.

This expansion makes it also easier to deal with the unwieldy
normalization factor $1/|\vec q\,|$ appearing in the covariant
representation~(\ref{polvector1}). In the hard region, we can express
it in terms of a power series in $1/N^2$ where
$N=(p_t+q)^2-m_t^2=2p_tq+q^2$ is the denominator of a top-quark
propagator with momentum $p_t+q$. Using again $p_t\cdot q=m_tq_0$ we
find 
\be\label{N}
4m_t^2|\vec q\,|^2=(N^2-2q^2N+q^4-4m_t^2q^2),
\en
which then leads to the expansion~\cite{Czarnecki:2010gb}
\begin{equation}\label{1/Nexpansion}
\frac1{|\vec q\,|}=\frac{2m_t}{N}\sum_{i=0}^\infty{2i\choose i}\!\!
  \left(\frac{2q^2N-q^4+4m_t^2 q^2}{4\,N^2}\right)^i.  
\end{equation}
In our calculation of the Feynman diagrams, we are only interested in
the imaginary part due to a cut through the $W$-boson line. Thus, we
can replace $q^2$ by $m_W^2$ in Eq.~(\ref{1/Nexpansion}). The series
is then truncated at the desired order in $x$.

In the soft region, it is not possible to construct an expansion of
$|\vec q\,|$, since $|\vec q\,|^2 = q_0^2 - m_W^2$ and $q_0\sim m_W$
in the soft region. However, in this region the loop containing the
$W$ boson factorizes from the remaining diagram due to the
expansion. Therefore, the only integrals that have to be modified are
one-loop massive tadpole integrals, which are relatively simple.

After the expansion, all remaining integrals depend only on a
single scale and are thus easier to compute. However, the denominators
of the expanded propagators are now raised to higher powers. We use
the program {\tt rows}~\cite{rows}, which implements the so-called
Laporta algorithm~\cite{Laporta:1996mq,Laporta:2001dd}, to reduce all
of these integrals to a small set of so-called master
integrals. Compared to the calculation of the unpolarized decay rate
and the helicity fractions, we do not encounter any new master
integrals.

%%%%%%%%%%%%%%%%%%%%%%%%%%%%%%%%%%%%%%%%%%%%%%%%%%%%%%%%%%%%%%%%%%%%%%%%%%%%%%%
\section{Numerical results\label{sec::num}}
%%%%%%%%%%%%%%%%%%%%%%%%%%%%%%%%%%%%%%%%%%%%%%%%%%%%%%%%%%%%%%%%%%%%%%%%%%%%%%%
Our analytical results can be found in the Appendix to this paper. For the
numerical evaluation of the analytical expression we use the values
$m_t= 173.1 \pm 0.6\GeV$,
$m_W=80.385\pm0.015\GeV$ and
$\alpha_s^{(5)}(m_Z)=0.1182\pm 0.0012$~\cite{Olive:2016xmw}. The
strong coupling constant is then evolved to the required scale using
five-loop running. Note that our result is expressed in terms of the
strong coupling constant with six active flavors, whereas the initial
value $\alpha_s^{(5)}(m_Z)$ is defined with only five. Thus, we also
have to use the (four-loop) decoupling relation to translate the
latter into the former. All of this is achieved with the help of
version 3 of the program
{\tt RunDec}~\cite{Chetyrkin:2000yt,Herren:2017osy}. Our central value
is $\alpha_s^{(6)}(m_t)=0.1078$.

We present our results in terms of the reduced helicity rates
$\hat\Gamma_\alpha$ defined by
\begin{equation}
\Gamma_\alpha=\frac{G_Fm_t^3|V_{tb}|^2}{8\sqrt2\pi}\,\hat\Gamma_\alpha.
\label{eq::redrate}
\end{equation}
The total unpolarized and polarized rates are denoted by $\alpha=U+L$ and
$\alpha=(U+L)^P$, where $L$ refers to the longitudinal and $U$ to the
unpolarized-transverse polarization of the $W$ boson (the latter is
the sum of the two transverse polarizations). We then expand the
reduced rates up to the second order in the strong coupling constant
$\alpha_s$ as
\begin{equation}
  \hat\Gamma_\alpha^{\mathrm{NNLO}}=\hat\Gamma_\alpha^{(0)}
  +\hat\Gamma_\alpha^{(1)}\pfrac{\alpha_s}{\pi}
  +\hat\Gamma_\alpha^{(2)}\pfrac{\alpha_s}{\pi}^2,
  \label{eq::pertseries}
\end{equation}
where $\alpha_s\equiv\alpha_s^{(6)}(m_t)$ is defined with six active
flavors and evaluated at the renormalization scale
$\mu=m_t$. Furthermore, we define the coefficients in the $x=m_W/m_t$
expansion by
\begin{equation}
  \hat\Gamma_{U+L}^{\mathrm{NNLO}}=
  \sum_{i=0}^{10}\hat\Gamma^{\mathrm{NNLO}}_i\, x^i,\qquad
  \hat\Gamma_{(U+L)^P}^{\mathrm{NNLO}}=
  \sum_{i=0}^{10}\hat\Gamma^{\mathrm{NNLO}}_{P,i}\, x^i.
  \label{eq::GamCoeff}
\end{equation}
For $\hat\Gamma^{(2)}_{U+L}$ we use the result of Ref.~\cite{Blokland:2005vq}.
Note that the coefficients of $\hat\Gamma_{U+L}$ contain
logarithms of $x$. In principle, the sums run up to infinity, but in
practice we only calculated the terms up to ${\cal O}(x^{10})$. This
is sufficient to provide a reliable approximation of the full
result. Up to the order ${\cal O}(x^n)$ we then calculate the
NLO and NNLO values of the asymmetry parameter according to the ratio
\begin{equation}
  \label{asymm}
  \alpha_P^{\mathrm{(N)NLO}}(n)=
  \frac{\sum_{i=0}^n\hat\Gamma^{\mathrm{(N)NLO}}_{P,i}\,  x^i}
  {\sum_{i=0}^n\hat\Gamma^{\mathrm{(N)NLO}}_{i} x^i}\,,
\end{equation}
where $\hat\Gamma_\alpha^{\mathrm{NLO}}$ is defined as in
Eq.~(\ref{eq::pertseries}), but with $\hat\Gamma_\alpha^{(2)}$ set to
zero.

\begin{table}[t]
\caption{\label{table1}Numerical values for coefficients in the $x$
  expansion of the unpolarized and polarized reduced rates
  $\hat\Gamma^{\mathrm{NNLO}}_{U+L}$ and $\hat\Gamma^{\mathrm{NNLO}}_{(U+L)^P}$
  (cf. Eq.~(\ref{eq::GamCoeff})). The results for the rates are given
  in the last line. In the third and fourth column we list the  values
  of the asymmetry parameter at NLO and NNLO at a given order $n$ in
  the $x$ expansion (cf. Eq.~(\ref{asymm})).}
\vspace{0.2cm}
\begin{center}
\begin{tabular}{l|llll}
  \hline
  \hline
$n$ &
$\hat \Gamma^{\mathrm{NNLO}}_{n}$ &
$\hat \Gamma^{\mathrm{NNLO}}_{P,n}$ &
$\alpha_P^{\mathrm{NLO}}(n)$&
$\alpha_P^{\mathrm{NNLO}}(n)$ \\ 
\hline
$0$  & $+0.88690$ & $+0.88360$ & $0.99671$ & $0.99628$ \\[0.5ex]
$2$  & $+0.07452$ & $-3.65421$ & $0.11502$ & $0.10582$ \\[0.5ex]
$4$  & $-2.93225$ & $+4.93143$ & $0.42533$ & $0.42381$ \\[0.5ex]
$5$  & $ 0$       & $-0.31636$ & $0.41792$ & $0.41490$ \\[0.5ex]
$6$  & $+2.02534$ & $-2.08447$ & $0.38221$ & $0.37763$ \\[0.5ex]
$7$  & $ 0$       & $+0.23471$ & $0.38338$ & $0.37901$ \\[0.5ex]
$8$  & $-0.15921$ & $-0.00614$ & $0.38351$ & $0.37916$ \\[0.5ex]
$9$  & $ 0$       & $+0.01129$ & $0.38352$ & $0.37918$ \\[0.5ex]
$10$ & $-0.03276$ & $-0.00048$ & $0.38352$ & $0.37919$ \\[0.5ex]
\hline
$\hat \Gamma^{\mathrm{NNLO}}_{\alpha}$ & $+0.78655$ & $+0.29825$ &  &  \\[0.1ex]
\hline
\hline
\end{tabular} 
\end{center}
\end{table}

\begin{figure}[t]
  \begin{center}
    \epsfig{figure=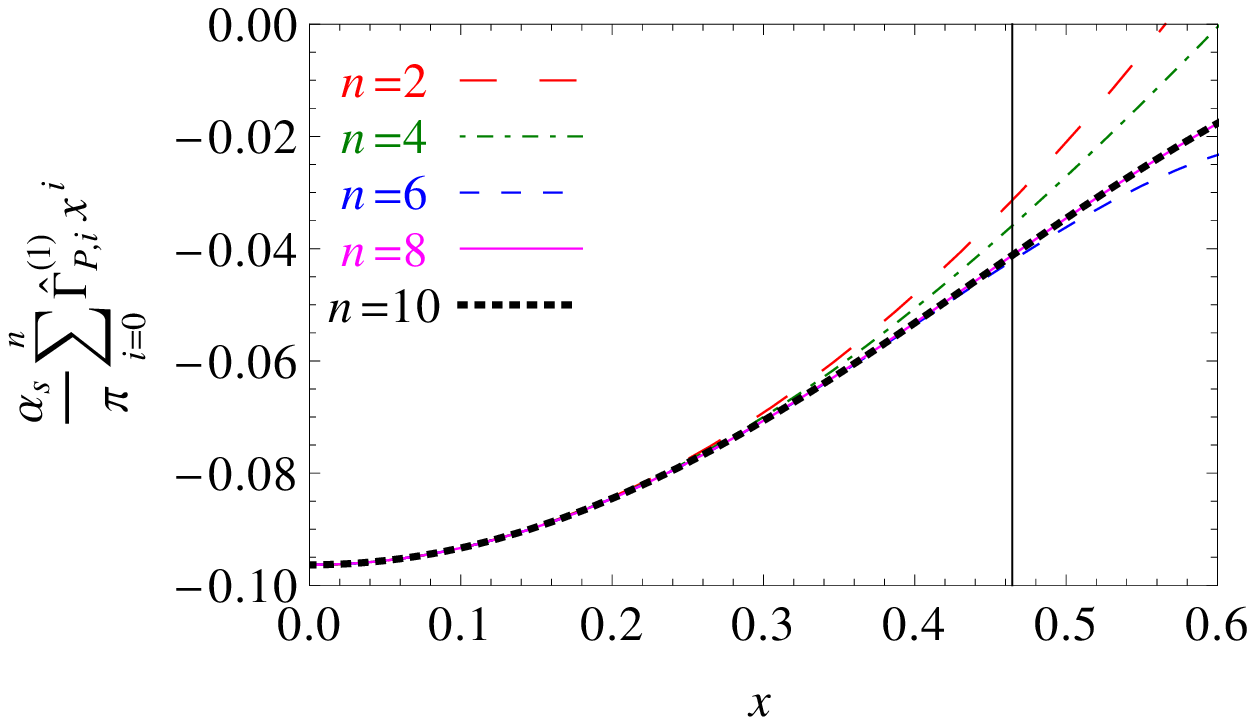, scale=0.8}\\
    \epsfig{figure=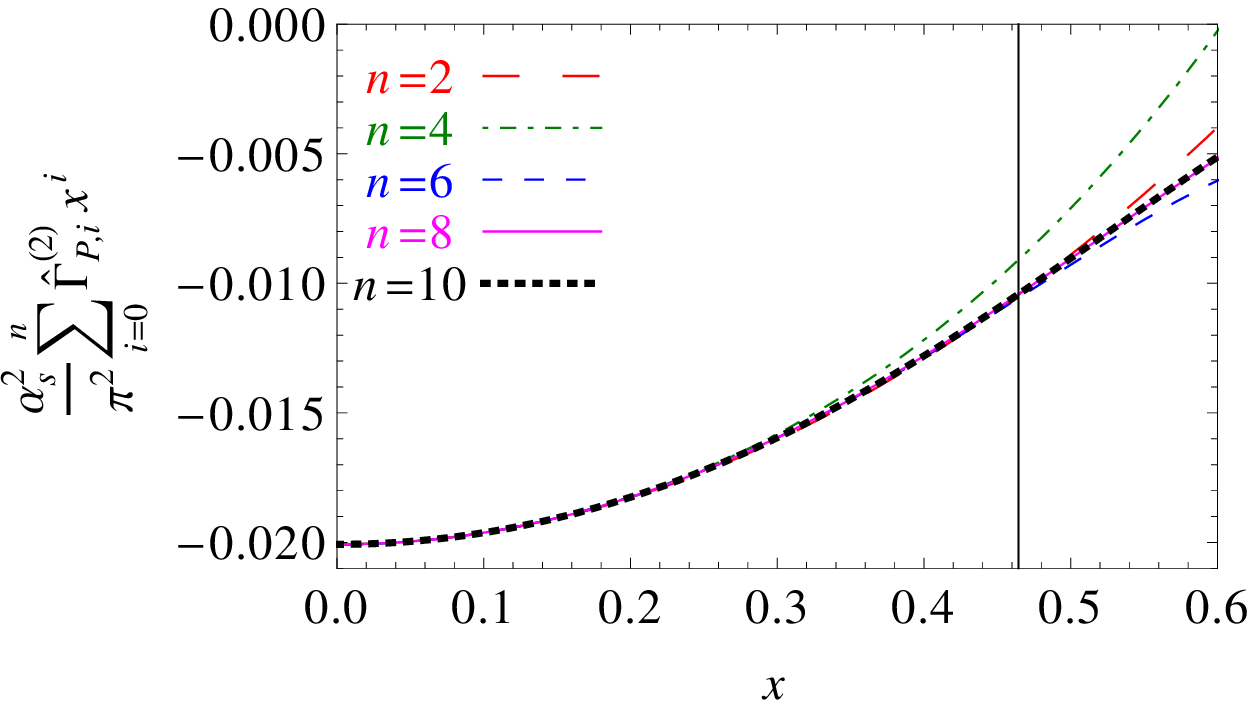, scale=0.8}
  \end{center}
  \caption{\label{fig::gam}The ${\cal O}(\alpha_s)$ and
    ${\cal O}(\alpha_s^2)$ contributions to
    $\hat\Gamma_{(U+L)^P}^{\mathrm{NNLO}}$ as functions of $x$. The
    coefficients $\hat\Gamma_{P,i}^{(k)}$ are defined analogously to
    the ones in Eq.~\ref{eq::GamCoeff}, but for
    $\hat\Gamma_{(U+L)^P}^{(k)}$ instead of
    $\hat\Gamma_{(U+L)^P}^{\mathrm{NNLO}}$. Our central value for
    $\alpha_s^{(6)}(m_t)$ is used throughout. The vertical line
    indicates the physical value of $x$. Note that the lines for $n=8$
    lie below the ones for $n=10$.}
\end{figure}

\begin{figure}[t]
  \begin{center}
    \epsfig{figure=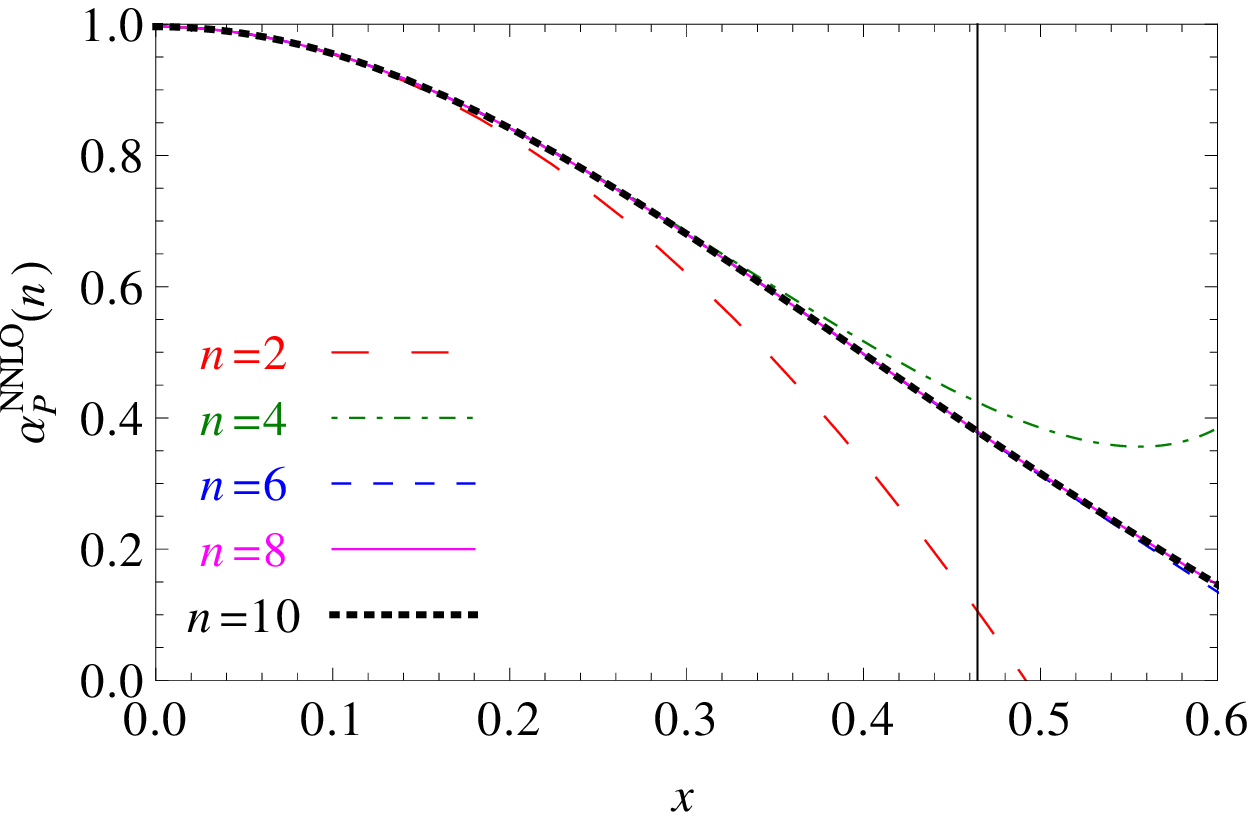, scale=0.8}
  \end{center}
  \caption{\label{fig::alp}$\alpha_P^{\mathrm{NNLO}}(n)$ as a function
    of $x$. Our central value for $\alpha_s^{(6)}(m_t)$ is used
    throughout. The vertical line indicates the physical value of
    $x$. Note that the lines for $n=6$ and $n=8$ lie below the one for
    $n=10$.}
\end{figure}

In Tab.~\ref{table1} we give numerical results for the coefficients of
the reduced rates and the asymmetry parameter. Analytical results are
given in Appendix~\ref{app::anares}. For the reduced rates, we find
that the absolute values of the coefficients in the power series in $x$
decrease when the power of $x$ increases. The convergence of the $x$
expansion is also illustrated in Figs.~\ref{fig::gam}
and~\ref{fig::alp}. Fig.~\ref{fig::gam} shows the ${\cal O}(\alpha_s)$
and ${\cal O}(\alpha_s^2)$ contributions to
$\hat\Gamma_{(U+L)^P}^{\mathrm{NNLO}}$ as functions of $x$. We observe
in both cases that adding terms beyond $x^6$ leads only to small
changes at the physical value of $x$. Furthermore, the results
truncated after $x^8$ and $x^{10}$ are visually indistinguishable even
up to $x=0.6$. A similar behavior can be observed for
$\alpha_P^{\mathrm{NNLO}}$ in Fig.~\ref{fig::alp}. (The figure for
$\alpha_P^{\mathrm{NLO}}$ would look very similar due to the smallness
of the ${\cal O}(\alpha_s^2)$ correction.) Finally, we note that the
unexpanded result for $\hat\Gamma_{(U+L)^P}^{(1)}$ given in the
Appendix would be indistinguishable from the $n=10$ curve in the upper
panel of Fig.~\ref{fig::gam}. Thus, we have good convergence behavior
and the truncation of the series does not change the result for all
practical purposes. Indeed, we can see in Tab.~\ref{table1} that the
difference between $\alpha^{\mathrm{(N)NLO}}_P(10)$ and
$\alpha^{\mathrm{(N)NLO}}_P(8)$ at the physical value for $x$ is
already at the level of $10^{-5}$.

In order to determine the precision of our final result for the
asymmetry parameter, we consider the following sources of
uncertainties:
\begin{itemize}
\item The uncertainty in the mass of the top quark. This is the
  largest source of uncertainty in our result. We note that in our
  calculation we have employed the pole mass definition for the top
  quark, whereas the numerical value corresponds to the so-called
  Monte-Carlo mass parameter. This difference adds an additional
  uncertainty to our result, which is, however, currently not
  precisely known and not included in our analysis. (Recent efforts to
  determine this difference can be found
  in Refs.~\cite{Butenschoen:2016lpz,Dehnadi:2018hrh}.)
\item Higher orders in QCD. We estimate the size of unknown higher
  order corrections by taking half the difference between
  $\alpha^{\mathrm{NNLO}}_P(10)$ and
  $\alpha^{\mathrm{NLO}}_P(10)$.\footnote{Alternatively, one could
    also vary the renormalization scale by a factor two around the
    central value $\mu=m_t$. This would give a value that is roughly
    one half of the one from our chosen method.}
\item The strong coupling constant. In addition to the uncertainty in
  the value of $\alpha_s^{(5)}(m_Z)$, we also vary the decoupling
  scale at which the five-flavor value is translated to the six-flavor
  one. However, the effect of the latter is completely negligible.
\item The uncertainty in the mass of the $W$ boson. 
\item The truncation of the series in $x$. We estimate this effect by
  taking the difference between $\alpha^{\mathrm{NNLO}}_P(10)$ and
  $\alpha^{\mathrm{NNLO}}_P(8)$. As can be seen from
  Tab.~\ref{table1}, this uncertainty is very small.
\item Non-zero bottom-quark mass. We estimate the error due to setting
  $m_b$ to zero by taking the difference to
  $\alpha^{\mathrm{NNLO}}_P(10)$ computed as before, but with
  $m_b=5\GeV$ in the Born-level contributions $\hat\Gamma_{U+L}^{(0)}$
  and  $\hat\Gamma_{(U+L)^P}^{(0)}$.
\end{itemize}

Our final result is
\begin{eqnarray}
  \alpha^{\mathrm{NNLO}}_P
  &=&
      0.3792
      \pm 0.0029\, (m_t)
      \pm 0.0022\, (\mbox{higher orders})
      \pm 0.0002\, (\alpha_s) \nonumber\\
  & &
      \phantom{0.3792}
      \pm 0.0002\, (m_W)
      \pm 0.00002\, (\mbox{truncation})
      \pm 0.0004\, (m_b\neq 0) \\
  &=&
      0.3792 \pm 0.0037 \,.
      \label{eq::apfinal}
\end{eqnarray}
In the last line, we have added the different uncertainties in
quadrature. It is important to note that the above result includes
only QCD corrections. However, at this level of precision, electroweak
corrections can also play a role. Since the electroweak NLO
corrections to $\hat\Gamma_{(U+L)^P}$ are currently unknown, we make
an estimate of their size
by looking at the known corrections to the helicity fractions, where
they increase the Born-level results by roughly
$2\%$~\cite{Do:2002ky}. The total decay rate is shifted by a similar
amount~\cite{Denner:1990ns,Eilam:1991iz}. Taking both of these
corrections into account changes our result for
$\alpha^{\mathrm{NNLO}}_P$ only at the permille level, which is well
within our uncertainty estimate.

%%%%%%%%%%%%%%%%%%%%%%%%%%%%%%%%%%%%%%%%%%%%%%%%%%%%%%%%%%%%%%%%%%%%%%%%%%%%%
\section{Summary and outlook\label{sec::sum}}
%%%%%%%%%%%%%%%%%%%%%%%%%%%%%%%%%%%%%%%%%%%%%%%%%%%%%%%%%%%%%%%%%%%%%%%%%%%%%
We have presented analytical and numerical results on the NNLO coefficients of
a power series expansion of the polarized decay rate where we have
expanded in the mass ratio $x=m_W/m_t$. Including the previously
calculated LO and NLO results and the NNLO result for the unpolarized
decay rate, we obtain a ${\cal O}(\alpha_s^2)$ result for the asymmetry
parameter $\alpha_P$ determining the angular decay distribution of a polarized
top quark decay.
It would be interesting to experimentally check on the size of the
asymmetry parameter in polarized top quark decays.

It is interesting to observe that the power series expansion of the parity-odd
polarized rate $\hat\Gamma_P^{(i)}$ contains both even and odd powers while
the parity-even unpolarized rate contains only even powers of $x$. This
follows the pattern observed in the NNLO calculation of the helicity
fractions~\cite{Czarnecki:2010gb}. We regret to say that we are lacking a deep
understanding of this pattern. We mention that the electroweak NLO corrections
to the structure functions do not follow this pattern.

In this paper we have summed over the three helicities of the $W$ boson. It
would be interesting to repeat the calculation for the three helicity
components of the $W$ boson separately. The corresponding decay distribution
is given by
\begin{eqnarray}
\frac1{\hat\Gamma}\,\frac{\mathrm{d}\hat\Gamma}{\mathrm{d}\cos\theta_P\,\mathrm{d}\cos\theta}
  &=&\frac12\Bigg\{\,\frac38\Big(1\!+\!\cos\theta \Big)^2
  (\hat\Gamma_++\hat\Gamma_+^P\,P_t\,\cos\theta_P\Big)\nn&&
  +\frac38\Big(1\!-\!\cos\theta\Big)^2
  (\hat\Gamma_-+\hat\Gamma_-^P\,P_t\,\cos\theta_P\Big)\nn&&
  +\frac34\sin^2\theta\Big(\hat\Gamma_L+\hat\Gamma_L^P\,P_t\,\cos\theta_P\Big)
  \Bigg\}.
\end{eqnarray}
It should be clear that all three asymmetries parameters
$\alpha^P_j=\hat\Gamma^P_j/\hat\Gamma_j$ ($j=+,-,L$) must satisfy the
positivity condition $|\alpha^P_j|\le 1$.

The LO Born term values for the unpolarized and polarized structure functions
are given by~\cite{Fischer:1998gsa,Fischer:2001gp}
\bea
\hat{\Gamma}_+ &=& 0, \qquad \qquad \qquad \quad \, \hat{\Gamma}_+^P=0, \nn
\hat{\Gamma}_- &=& 2x^2(1-x^2)^2, \qquad \,\,\hat{\Gamma}_-^P=-2x^2(1-x^2)^2, \nn
\hat{\Gamma}_L &=& (1-x^2)^2, \qquad \qquad \hat{\Gamma}_L^P=(1-x^2)^2.
\ena
The LO asymmetry parameter $\alpha^P_j$ is undetermined for the transverse-plus
rate and maximal for the transverse-minus and the longitudinal rate. This has
to be compared to the total LO asymmetry parameter $\alpha_P^{\mathrm{LO}}= 0.398$ which is
far from being maximal.

Including the NLO corrections one obtains $|\alpha^{P,\mathrm{NLO}}_j|<1$ for all
three asymmetry parameters~\cite{Fischer:2001gp}. This is very gratifying from
the point of view that the ${\cal O}(\alpha_s)$ asymmetry parameters satisfy the
necessary positivity condition $|\alpha^P_j|\le1$. We expect that the inclusion
of NNLO results in the calculation of the asymmetry parameter will retain this
feature.

%%%%%%%%%%%%%%%%%%%%%%%%%%%%%%%%%%%%%%%%%%%%%%%%%%%%%%%%%%%%%%%%%%%%%%%%%%%%
%%%%%%%%%%%%%%%%%%%%%%%%%%%%%%%%%%%%%%%%%%%%%%%%%%%%%%%%%%%%%%%%%%%%%%%%%%%%
\subsection*{Acknowledgments}
We would like to thank J. Mueller for encouragement.
The loop diagrams were calculated with {\tt FORM}~\cite{Vermaseren:2000nd}.
This work was supported by the Estonian Science Foundation under grant
No.~IUT2-27.
A.C.~was supported by the Natural Sciences and Engineering Research Council of Canada.
S.G.\ acknowledges the support of the theory group THEP at the
Institute of Physics at the University of Mainz and the support of the Cluster
of Excellence PRISMA at the University of Mainz.
 
%%%%%%%%%%%%%%%%%%%%%%%%%%%%%%%%%%%%%%%%%%%%%%%%%%%%%%%%%%%%%%%%%%%%%%%%%%%%%%
\appendix
\section{Analytical results\label{app::anares}}
\setcounter{equation}{0}\def\theequation{A\arabic{equation}}
%%%%%%%%%%%%%%%%%%%%%%%%%%%%%%%%%%%%%%%%%%%%%%%%%%%%%%%%%%%%%%%%%%%%%%%%%%%%%%
In this Appendix we provide the analytical results for the reduced
rates defined in Eqs.~(\ref{eq::redrate},\ref{eq::pertseries}).

\subsubsection*{LO Born term contributions}
\begin{eqnarray}
\hat\Gamma_{U+L}^{(0)}&=&(1-x^2)^2(1+2x^2),\nonumber\\
\hat\Gamma_{(U+L)^P}^{(0)}&=&(1-x^2)^2(1-2x^2).
\end{eqnarray}
\subsubsection*{\boldmath NLO $\alpha_s$-corrections}
Using the techniques described in the main part of the paper we calculate the
NLO corrections in $\alpha_s$ in terms of a series expansion in the mass ratio
$x=m_W/m_t$. One has
\begin{eqnarray}
\hat\Gamma_{U+L}^{(1)}&=&C_F\Bigg[\frac54+\frac32x^2-6x^4
  +\frac{46}9x^6-\frac74x^8-\frac{49}{300}x^{10}\strut\nonumber\\&&\strut\qquad
  -2(1-x^2)^2(1+2x^2)\zeta(2)
  +\left(3-\frac43x^2+\frac32x^4+\frac25x^6\right)x^4\ln x\Bigg]\,,\nonumber\\
\hat\Gamma_{(U+L)^P}^{(1)}&=&C_F\Bigg[-\frac{15}4-\frac{17}8x^4
  -\frac{1324}{225}x^5-\frac{31}{36}x^6\strut\nonumber\\&&\strut\qquad
  +\frac{48868}{11025}x^7-\frac{23}{288}x^8+\frac{884}{6615}x^9
  -\frac3{100}x^{10}+(1+4x^2)\zeta(2)\Bigg]\,,
\end{eqnarray}
where $C_F=(N_c^2-1)/(2N_c)=4/3$ for $N_c=3$ colors and $\zeta$
denotes the Riemann zeta function. These results can be
compared with the $x$ expansion of the closed form results calculated in
Refs.~\cite{Fischer:1998gsa,Fischer:2001gp}. One has
\begin{eqnarray}
\hat\Gamma_{U+L}^{(1)}&=&C_F
  \Bigg[\frac14(1-x^2)(5+9x^2-6x^4)-2x^2(1+x^2)(1-2x^2)\ln x
  \strut\nonumber\\&&\strut
  -\frac12(1-x^2)^2(5+4x^2)\ln(1-x^2)\strut\nonumber\\&&\strut
  -2(1-x^2)^2(1+2x^2)\left(2\Li_2(x)+2\Li_2(-x)+\ln x\ln(1-x^2)
    +\frac{\pi^2}6\right)\Bigg],\nonumber\\
\hat\Gamma_{(U+L)^P}^{(1)}&=&C_F\Bigg[
  -\frac14(1-x)^2(15+2x-5x^2-12x^3+2x^4)+(1+4x^2)\zeta(2)
  \strut\nonumber\\&&\strut
  -\frac12(1-x^2)^2(1-4x^2)\ln(1-x)-\frac12(1-x^2)(3-x^2)(1+4x^2)\ln(1+x)
  \strut\nonumber\\&&\strut
  -2(1-x^2)^2(1-2x^2)\Li_2(x)+2(2+5x^4-2x^6)\Li_2(-x)\Bigg],
\end{eqnarray}
where $\Li_2$ denotes the dilogarithm function.
We have found agreement in this comparison.

\subsubsection*{\boldmath NNLO $\alpha_s^2$-corrections}
We present our results in terms of the color-flavor decomposition
\begin{equation}
\hat\Gamma_\alpha^{(2)}=C_F\left[C_F\hat\Gamma_\alpha^{(2F)}
  +C_A\hat\Gamma_\alpha^{(2A)}+N_LT_F\hat\Gamma_\alpha^{(2L)}
  +N_HT_F\hat\Gamma_\alpha^{(2H)}\right],
\end{equation}
where $C_A=N_c=3$, $T_F=1/2$, $N_L=5$ and $N_H=1$.
The coefficients of $\hat\Gamma^{(2)}_{U+L}$ were calculated in
Ref.~\cite{Blokland:2005vq} and are presented here for completeness.
We have
\begin{eqnarray}
\lefteqn{\hat\Gamma_{U+L}^{(2F)}\ =\ 5-\frac{73}8x^2
  -\frac{7537}{288}x^4+\frac{16499}{864}x^6-\frac{1586479}{259200}x^8
  -\frac{11808733}{6480000}x^{10}\strut}\nonumber\\&&\strut
  +\left(\frac{115}{24}-\frac{367}{72}x^2+\frac{31979}{8640}x^4
    +\frac{13589}{13500}x^6\right)x^4\ln x\strut\nonumber\\&&\strut
  -\Bigg(\frac{119}8-\frac{123}4x^2-\frac{523}{16}x^4+\frac{407}{36}x^6
    -\frac{2951}{1152}x^8-\frac{37}{400}x^{10}\strut\nonumber\\&&\strut\qquad
    -\left(\frac{57}2-\frac{81}8x^4-6x^6\right)\ln 2
    +\left(\frac{15}4-\frac{20}3x^2+\frac34x^4+\frac15x^6\right)
    x^4\ln x\Bigg)\zeta(2)\strut\nonumber\\&&\strut
  -\left(\frac{53}8-\frac{295}{32}x^4+\frac72x^6-\frac92x^8-\frac65x^{10}
    \right)\zeta(3)
  -\left(\frac{11}8+41x^2+\frac{191}8x^4-\frac{21}4x^6\right)\zeta(4),
  \nonumber\\[12pt]
\lefteqn{\hat\Gamma_{(U+L)^P}^{(2F)}\ =\ -\frac{35}{48}
  -\frac{3245}{48}x^2+\frac{132413}{11520}x^4-\frac{6991909}{405000}x^5
  +\frac{1931557}{72576}x^6\strut}\nonumber\\&&\strut\qquad
  +\frac{13210017881}{972405000}x^7-\frac{68041043843}{1219276800}x^8
  +\frac{92602080451}{35006580000}x^9-\frac{4454582599}{14515200}x^{10}
  \strut\nonumber\\&&\strut
  -\Bigg(\frac{35}4-61x^2-\frac{1889}{32}x^4-\frac{862}{75}x^5
    -\frac{4529}{72}x^6+\frac{87146}{11025}x^7\strut\nonumber\\&&\strut\qquad
    -\frac{1674161}{9216}x^8+\frac{31246}{19845}x^9
    -\frac{122414357}{230400}x^{10}\strut\nonumber\\&&\strut\qquad
    -\left(\frac{55}2-19x^2+\frac{93}8x^4-\frac{1279}{16}x^6
      -\frac{64787}{256}x^8-\frac{24113}{32}x^{10}\right)\ln 2\Bigg)\zeta(2)
  \strut\nonumber\\&&\strut
  -\left(\frac{95}8-\frac{113}4x^2+\frac{5927}{160}x^4-\frac{70097}{1344}x^6
    -\frac{11855441}{107520}x^8-\frac{286453}{896}x^{10}\right)\zeta(3)
  \strut\nonumber\\&&\strut
  -\left(\frac38+\frac{177}4x^2+\frac{605}8x^4+\frac{337}4x^6
    +\frac{171}2x^8+\frac{171}2x^{10}\right)\zeta(4)\strut\nonumber\\&&\strut
  -4(1-2x^2)(2-2x^2+x^4)\left(\Li_4\pfrac12-\ln^22\zeta(2)+\frac1{24}\ln^42
  \right),\nonumber\\[12pt]
\lefteqn{\hat\Gamma_{U+L}^{2A}\ =\ \frac{521}{576}
  +\frac{91}{48}x^2-\frac{12169}{576}x^4+\frac{13685}{864}x^6
  -\frac{420749}{103680}x^8-\frac{4868261}{12960000}x^{10}
  \strut}\nonumber\\&&\strut
  +\left(\frac{73}8-\frac{1121}{216}x^2+\frac{11941}{3456}x^4
    +\frac{153397}{108000}x^6\right)x^4\ln x\strut\nonumber\\&&\strut
  +\Bigg(\frac{505}{144}+\frac{329}{24}x^2+\frac{2171}{96}x^4-\frac{47}{12}x^6
  -\frac{3263}{2304}x^8-\frac{557}{800}x^{10}\strut\nonumber\\&&\strut\qquad
  -\left(\frac{57}4-\frac{81}{16}x^4-3x^6\right)\ln 2
  -\left(\frac98+2x^2+\frac98x^4+\frac3{10}x^6\right)x^4\ln x\Bigg)\zeta(2)
  \strut\nonumber\\&&\strut
  +\left(\frac9{16}+\frac{377}{64}x^4-\frac{19}4x^6-\frac98x^8
  -\frac3{10}x^{10}\right)\zeta(3)
  +\left(\frac{11}{16}-\frac{39}2x^2-\frac{385}{16}x^4+\frac{43}8x^6\right)
  \zeta(4),\nonumber\\[12pt]
\lefteqn{\hat\Gamma_{(U+L)^P}^{2A}\ =\ -\frac{3155}{192}
  +\frac{15}{16}x^2-\frac{5213}{384}x^4-\frac{645811}{40500}x^5
  +\frac{6888169}{259200}x^6\strut}\nonumber\\&&\strut\qquad
  +\frac{19545586}{1929375}x^7+\frac{7008567101}{101606400}x^8
  +\frac{8723471549}{26254935000}x^9+\frac{117991469621}{609638400}x^{10}
  \strut\nonumber\\&&\strut
  +\Bigg(\frac{1129}{144}+\frac{455}{18}x^2+\frac{3229}{192}x^4
  +\frac{31}{225}x^5+\frac{373}{144}x^6\strut\nonumber\\&&\strut\qquad
  -\frac{353}{735}x^7-\frac{345851}{6144}x^8+\frac{1853}{2835}x^9
  -\frac{35471879}{153600}x^{10}\strut\nonumber\\&&\strut\qquad
  -\left(\frac{55}4-\frac{19}2x^2+\frac{93}{16}x^4-\frac{1279}{32}x^6
  -\frac{64787}{512}x^8-\frac{24113}{64}x^{10}\right)\ln 2\Bigg)\zeta(2)
  \strut\nonumber\\&&\strut
  +\left(\frac{191}{16}-\frac{427}{24}x^2+\frac{6329}{192}x^4
  -\frac{4909}{128}x^6-\frac{12031441}{215040}x^8-\frac{4301531}{26880}x^{10}
  \right)\zeta(3)\strut\nonumber\\&&\strut
  -\left(\frac94+\frac{117}8x^2+\frac{531}{16}x^4+\frac{373}8x^6
  +\frac{93}2x^8+\frac{93}2x^{10}\right)\zeta(4)\strut\nonumber\\&&\strut
  +2(1-2x^2)(2-2x^2+x^4)\left(\Li_4\pfrac12-\ln^22\zeta(2)+\frac1{24}\ln^42
  \right),\nonumber\\[12pt]
\lefteqn{\hat\Gamma_{U+L}^{2L}\ =\ -\frac49-\frac{19}6x^2
  +\frac{745}{72}x^4-\frac{5839}{648}x^6+\frac{4253}{8640}x^8
  -\frac{689}{27000}x^{10}\strut}\nonumber\\&&\strut
  -\left(\frac72-\frac{10}3x^2+\frac{17}{72}x^4+\frac7{450}x^6\right)x^4\ln x
  \strut\nonumber\\&&\strut
  +\left(\frac{23}{18}+\frac43x^2-\frac{31}6x^4+\frac{14}9x^6+\frac32x^8
  +\frac25x^{10}\right)\zeta(2)+(1-x^2)^2(1+2x^2)\zeta(3),\nonumber\\[12pt]
\lefteqn{\hat\Gamma_{(U+L)^P}^{2L}\ =\ \frac{19}4-\frac12x^2
  +\frac{1565}{288}x^4+\frac{20249}{3375}x^5+\frac{9319}{6480}x^6
  \strut}\nonumber\\&&\strut\qquad
  -\frac{437779}{128625}x^7-\frac{513487}{725760}x^8
  -\frac{4993343}{12502350}x^9-\frac{284003}{2268000}x^{10}
  \strut\nonumber\\&&\strut
  +\left(\frac{10}9-\frac{44}9x^2+\frac56x^4+\frac89x^6-\frac7{18}x^8
  -\frac4{45}x^{10}\right)\zeta(2)
  -\left(3-\frac43x^2+\frac{25}3x^4-\frac{10}3x^6\right)\zeta(3),
  \nonumber\\[12pt]
\lefteqn{\hat\Gamma_{U+L}^{2H}\ =\ \frac{12991}{1296}
  -\frac{35}{108}x^2-\frac{6377}{432}x^4+\frac{319}{27}x^6
  +\frac{76873}{8640}x^8+\frac{237107}{27000}x^{10}\strut}\nonumber\\&&\strut
  -\left(\frac{53}9+\frac83x^2-\frac{25}3x^4+\frac{62}9x^6+\frac{16}3x^8
  +\frac{16}3x^{10}\right)\zeta(2)
  -\left(\frac13-4x^2-x^4+\frac23x^6\right)\zeta(3),
\nonumber\\[12pt]
\lefteqn{\hat\Gamma_{(U+L)^P}^{2H}\ =\ \frac{12991}{1296}
  -\frac{79}{81}x^2+\frac{15197}{1296}x^4-\frac{5005}{324}x^6
  -\frac{328061}{25920}x^8-\frac{2032763}{162000}x^{10}
  \strut}\nonumber\\&&\strut
  -\left(\frac{53}9+\frac{28}9x^2+\frac{109}9x^4-\frac{10}3x^6
  -\frac{16}9x^8-\frac{16}9x^{10}\right)\zeta(2)\strut\nonumber\\&&\strut
  -\left(\frac13-\frac{16}3x^2-\frac{19}3x^4-\frac{26}3x^6-8x^8-8x^{10}
  \right)\zeta(3),
\end{eqnarray}
where $\Li_4\pfrac12 = \sum\limits_{i=1}^\infty \frac{1}{2^n n^4}$.

\end{document}